\documentclass{pasj00}

\begin{document}
\SetRunningHead{Takakuwa et al.}{ASTE Observations of Low-mass Protostellar Envelopes}
\Received{2006/05/22}
\Accepted{2006/07/29}

\title{ASTE Observations of Warm Gas\\in Low-mass Protostellar Envelopes:\\
Different Kinematics\\between Submillimeter and Millimeter Lines}

\author{Shigehisa \textsc{Takakuwa}\thanks{e-mail: s.takakuwa@nao.ac.jp},
Takeshi \textsc{Kamazaki}, Masao \textsc{Saito}}
\affil{ALMA Project Office, National Astronomical Observatory of Japan,\\
Osawa 2-21-1, Mitaka, Tokyo, 181-8588, Japan}
\author{Nobuyuki \textsc{Yamaguchi}}
\affil{Nobeyama Radio Observatory, National Astronomical Observatory of Japan,\\
Nobeyama, Minamimaki, Minamisaku, Nagano, 384-1305, Japan}
\and
\author{Kotaro {\sc Kohno}}
\affil{Institute of Astronomy, The University of Tokyo, Osawa 2-21-1, Mitaka, Tokyo, 
181-0015, Japan}

%


\maketitle

\begin{abstract}
With the ASTE telescope, we have made observations of three low-mass protostellar
envelopes around L483, B335, and L723 in the submillimeter CS ($J$=7--6) and
HCN ($J$=4--3) lines.
We detected both the CS and HCN lines toward all the targets, and the typical
CS intensity ($\sim$ 1.0 K in T$_{B}$) is twice higher than that of the HCN line.
Mapping observations of L483 in these lines have shown that the
submillimeter emissions in the low-mass protostellar envelope are resolved,
exhibit a western extension from the central protostar, and that the deconvolved size
is $\sim$ 5500 AU $\times$ 3700 AU (P.A. = 78$^{\circ}$) in the HCN emission.
The extent of the submillimeter emissions in L483 implies the presence
of higher-temperature ($\gtrsim$ 40 K) gas at 4000 AU away from the central protostar,
which suggests that we need to take 2-dimensional radiative transfer models with
a flattened disklike envelope and bipolar cavity into account to
explain the temperature structure inside the low-mass protostellar envelope.
The position-velocity diagrams of these submillimeter lines in L483 and B335
exhibit different velocity gradients from those found in the previous millimeter
observations. In particular, along the axis of the associated molecular outflow
the sense of the velocity gradient traced by the submillimeter lines
is opposite to that of the millimeter observations or
the associated molecular outflow, both in L483 and B335.
We suggest that expanding
gas motions at the surface of the flattened disklike envelope around the protostar,
which is irradiated from the central star directly,
are the origin of the observed submillimeter velocity structure.
\end{abstract}

\section{Introduction}

Previous millimeter molecular-line observations have found 3000 -- 10000-AU-scale
molecular envelopes with both rotation and infalling gas motion around low-mass protostars
(\cite{oh96b,oh97a,oh97b,ta03b}).
These millimeter observations have probed structures and kinematics of molecular gas
with a temperature of $\sim$ 10 K and a density of $\sim$ 10$^{4-5}$ cm$^{-3}$
in the envelopes (\cite{tak00,sai01}).
However, structures and kinematics of higher-temperature ($\gtrsim$ 40 K) or
higher-density gas ($\gtrsim$ 6 $\times$ 10$^{6}$ cm$^{-3}$) in those envelopes,
which are presumably related to the region in the vicinity
of the protostar and disk system (\cite{cec00,oso03}), are poorly known.
This is mainly because millimeter observations can not trace
warm ($\gtrsim$ 40 K) and dense ($\gtrsim$ 6 $\times$ 10$^{6}$ cm$^{-3}$)
regions of envelopes due to the contamination from 
extended low-density ($\sim$ 10$^{4-5}$ cm$^{-3}$) and cold ($\sim$ 10 K) gas.
Submillimeter observations, which can trace warm ($\gtrsim$ 40 K) and dense 
($\gtrsim$ 6 $\times$ 10$^{6}$ cm$^{-3}$) gas selectively, are required to investigate such regions and 
to study the kinematics.

For these purposes, we have made Submillimeter Array (SMA)
\footnote{The Submillimeter Array (SMA) is a joint project between the
Smithsonian Astrophysical Observatory and the Academia Sinica Institute of
Astronomy and Astrophysics, and is funded by the Smithsonian Institution
and the Academia Sinica.} observations of
low-mass protostellar envelopes around L1551 IRS5 and IRAS 16293-2422
in the CS ($J$ = 7--6; 342.9 GHz) and HCN ($J$ = 4--3; 354.5 GHz) lines, respectively
(\cite{tak04,tak06}).
These submillimeter molecular lines trace gas temperatures above $\gtrsim$ 40 (K) and 
densities above $\gtrsim$ 6 $\times$ 10$^{6}$ cm$^{-3}$, and hence are appropriate tracers 
for our study.
Both in L1551 IRS5 and IRAS 16293-2422, we have detected a compact
($\sim$ 500 AU) disklike structure associated with the protostars in the
submillimeter lines.
However, our SMA data recover only $\sim$ 11$\%$ and $\sim$ 30$\%$ of the total
flux observed with JCMT in L1551 IRS5 and IRAS 16293-2422, respectively,
which implies that there is also an extended ($>$ 2000 AU) submillimeter component
in the low-mass protostellar envelopes.
In fact,
our JCMT mapping observations of IRAS 16293-2422 in the HCN (4--3) emission
have revealed an extended ($\sim$ 3000 AU) envelope structure as well as the compact 
($\sim$ 500 AU) disklike structure which is embedded in the extended envelope.
These results suggest that there is extended ($>$ 2000 AU) warm ($\gtrsim$ 40 K) or dense
($\gtrsim$ 6 $\times$ 10$^{6}$ cm$^{-3}$) gas 
as well as the compact disklike structure in these low-mass protostellar envelopes.
It seems to be difficult to make gas temperatures high enough strictly via heating
from the central stars embedded in the envelopes (e.g., \cite{lay94}),
and it is a puzzle why such
warm ($\gtrsim$ 40 K) or dense ($\gtrsim$ 6 $\times$ 10$^{6}$ cm$^{-3}$) gas
can be so extended in the low-mass protostellar envelopes.
These results suggest that we need to re-consider the origin of submillimeter emissions
in low-mass protostellar envelopes.

In order to clarify the extent, kinematics, and the origin of submillimeter emissions
in low-mass protostellar envelopes unambiguously, we have initiated mapping observations
of low-mass protostellar envelopes in the HCN (4--3) and CS (7--6) lines with ASTE
(Atacama Submillimeter Telescope Experiment),
a new and powerful submillimeter single-dish telescope at the Atacama Site in Chile.
In this paper, we report ASTE results of HCN (4--3) and CS (7--6) observations of
three nearby ($\leq$ 300 pc)
low-mass protostellar sources, that is, L483, B335, and L723.
L483 is a bright ($L_{bol}$ $\sim$ 14 $L_{\odot}$)
Class 0 object with an East (red) - West (blue) molecular outflow
at a distance of 200 pc (\cite{lad91,hat99,taf00}).
Millimeter CS (2--1) and H$_{2}$CO (2$_{12}$--1$_{11}$, 3$_{12}$--2$_{11}$) spectra
toward the protostar of L483 exhibit the "infalling asymmetry"
with the stronger blueshifted emission and the absorption dip,
and this source is considered to be one of the protostellar candidates with an infalling gas motion 
in the surrounding envelope (\cite{mye95,mar97}).
Mapping observations of L483 in centimeter and millimeter lines such
as NH$_{3}$ (1,1; 2,2), N$_{2}$H$^{+}$ (1--0) and CH$_{3}$OH (2--1) have revealed that
the protostellar envelope is extended along the east-west direction,
and that the velocity structure is similar to that of the associated molecular outflow,
suggesting that the dense gas in the outer portion of the envelope is being swept-up by the
associated outflow (\cite{gre97,ful00,par00,taf00}).
In L483, detailed studies of chemical differentiation in the envelope
have also been performed (\cite{jor02,jor04}).
B335 is a close ($d$ $\sim$ 250 pc) Bok globule associated with a far-infrared source
(IRAS 19347+0727; $L_{bol}$ $\sim$ 3 $L_{\odot}$),
and its spectral energy distribution classifies this source as a Class 0 object
(\cite{dav87,bar95}).
This source has an East (blue) - West (red) molecular outflow (\cite{hir92,cha93}).
Aperture synthesis observations of the millimeter H$^{13}$CO$^{+}$ emission (\cite{sai99}),
as well as the shape of the central spectra in the CS (2--1 ,3--2, 5--4) lines
(\cite{cho95,zho95,eva05}),
have shown that there is an infalling gas motion in the surrounding envelope around B335,
although the CS results may be severely affected by the contamination
from the molecular outflow (\cite{wil00}).
Detailed millimeter continuum studies in the context of the circumstellar disk and envelope
have also been performed in B335 (\cite{ha03a,ha03b}).
L723 is another Class 0 candidate with a luminosity of $\sim$ 3$L_{\odot}$
at a distance of 300 pc (\cite{dav87}).
In L723, there are two radio sources, VLA1 and VLA2
(\cite{ang91}), where only VLA2 is associated with the surrounding molecular envelope
as well as the quadrupolar CO outflow, suggesting that VLA2 is younger than VLA1
(\cite{hay91,hir98}).

We note that only limited single-dish mapping observations
of low-mass protostellar envelopes in submillimeter molecular lines have
been reported (\cite{hog98,hog99}), and that most submillimeter single-dish studies of low-mass 
protostellar envelopes are single-point observations, on the assumption that
submillimeter lines in
low-mass protostellar envelopes are more compact than the beam size of the single-dish
telescopes ($<$ 20$\arcsec$ $\sim$ 4000 AU in nearby sources) (\cite{bla94,mor95,dis95}).
Our new high-sensitivity mapping observations of the low-mass
protostellar envelopes in the submillimeter lines with ASTE provide us a
new insight of the origin of the submillimeter emission in low-mass protostellar
envelopes, as will be shown in the present paper.

\section{Observations}

With the ASTE 10-m telescope we made HCN ($J$ = 4--3; 354.505480 GHz)
and CS ($J$ = 7--6; 342.882866 GHz) observations of L483, B335, and L723
on August 16-20, 2005. Details of the ASTE telescope are presented by
Ezawa et al. (2004).
Table \ref{tab:obs} summarizes the observational parameters.
A cartridge-type 350 GHz receiver mounted on ASTE
is a double sideband instrument with an IF frequency range 
of 5 to 7 GHz (\cite{koh05}), and the HCN and CS lines were observed
simultaneously at different sidebands. In L483 and B335,
we made mapping observations centered on the protostellar positions 
at a grid spacing of 10$\arcsec$, providing Nyquist-sampled maps.
The total number of observed points are 34 and 21 in L483 and B335, respectively, 
which covers most of the line-emitting regions.
In L723 we only made a single-point observation toward the protostellar position.
The telescope pointing was checked at every $\sim$ 2 hours by making
continuum observations of Jupiter and Uranus,
and was found to be better than 2$\arcsec$ during the whole observing period.
As a standard source we also observed IRAS 16293-2422 after each pointing run,
and confirmed that the relative intensity is consistent within $\sim$ 40$\%$.
We compared the observed HCN and CS spectra toward IRAS 16293-2422
to those obtained with the CSO telescope which has the same dish size as that of the ASTE
telescope (\cite{bla94,dis95}), and found that the main beam efficiency of
the ASTE telescope was $\sim$ 0.6. We also confirmed that the asymmetric spectral shapes
in IRAS 16293-2422 taken with the ASTE telescope
are in excellent agreement with those taken with the CSO telescope.
Hereafter we show observed line intensities in the unit of T$_{B}$.

\begin{table}
  \caption{Parameters of the ASTE Observations}\label{tab:obs}
  \begin{center}
    \begin{tabular}{lcccc}
      \hline\hline
         Parameter &\multicolumn{4}{c}{Value} \\
       \hline
         Observing Date &\multicolumn{4}{c}{August 16-20, 2005} \\
         Target                 &L483   &B335  &L723  &IRAS 16293-2422 (Calibrator) \\
         Right Ascension (J2000) &18$^{h}$17$^{m}$29$^{s}$.86           &19$^{h}$37$^{m}$00$^{s}$.94          &19$^{h}$17$^{m}$53$^{s}$.62          &16$^{h}$32$^{m}$22$^{s}$.87  \\
         Declination (J2000)     &-04$^{\circ}$39$\arcmin$38$\arcsec$.7 &07$^{\circ}$34$\arcmin$08$\arcsec$.8 &19$^{\circ}$12$\arcmin$19$\arcsec$.5 &-24$^{\circ}$28$\arcmin$36$\arcsec$.6  \\
	 Distance to the Target       &200 pc   &250 pc & 300 pc  &160 pc \\
         Molecular Line &\multicolumn{4}{c}{HCN ($J$=4--3; 354.50548 GHz) $\&$ CS ($J$=7--6; 342.882866 GHz)} \\     
         Spatial  Resolution  &\multicolumn{4}{c}{$\sim$ 22$\arcsec$} \\
         Spectral Resolution  &\multicolumn{4}{c}{125 kHz ($\sim$ 0.11 km s$^{-1}$) } \\
         Main Beam Efficiency &\multicolumn{4}{c}{$\sim$ 60$\%$} \\
         Integration Time (center)   &15.0 min      &15.5 min       &23.7 min &$\sim$ 3 min at each run\\
         Integration Time (mapping)  &$\sim$ 10 min &$\sim$ 7.5 min &-  &- \\
         rms (mapping)                     &\multicolumn{2}{c}{$\sim$ 0.13 K in T$_{B}$}  &-  &-  \\
         Typical System Temperature (DSB)  &\multicolumn{4}{c}{200 - 500 K} \\
         Pointing Error  &\multicolumn{4}{c}{$<$ 2$\arcsec$} \\
         Pointing Source &\multicolumn{4}{c}{Jupiter, Uranus} \\
       \hline
    \end{tabular}

  \end{center}
\end{table}

\section{Results}
\subsection{HCN and CS Spectra}

In Figure \ref{fig:f1}, we show HCN (4--3) and CS (7--6) spectra toward the
protostellar positions of L483, B335, and L723. We succeeded to detect
these submillimeter lines in all the three low-mass protostars. 
In Table \ref{tab:lines}, we summarize line parameters of these submillimeter spectra
derived from the Gaussian fittings.
At the central protostellar positions,
intensities of the CS line range from 0.8 to 1.8 K
while intensities of the HCN line are more than a factor of two weaker (0.3 - 0.8 K).
The CS-to-HCN intensity ratio is roughly constant (2.1 - 2.5) across the three samples,
and there may be a slight correlation between the source luminosity and the line
intensities.
On the other hand, line widths of the HCN line are almost twice wider than
the CS line widths in L483 and L723, but much narrower in B335.
The HCN (4--3) line consists of six hyperfine components, two of which
($F$ = 3--3 and 4--4) can be separated from the main line ($F$ = 4--3)
in our data by 1.977 MHz (-1.67 km s$^{-1}$) and -1.610 MHz (1.36 km s$^{-1}$), respectively
(\cite{jew97}). The intensity ratio
between the main and the other two hyperfine lines at the local thermal equilibrium
condition is 0.0217 (\cite{jew97}), and these weaker hyperfine lines are unlikely to be detectable
with the present observations. However, it is still possible that the broader HCN
line widths found in L483 and L723 are due to the presence of the hyperfine lines,
if there is any hyperfine anomaly and the intensities of the $F$ = 3--3 and 4--4 lines
are significantly larger than the statistical values in L483 and L723.
As will be shown in the next section, the velocity structure traced by the HCN line
is consistent with that traced by the CS (7--6) line without any hyperfine component,
so the presence of the hyperfine components in the HCN line is unlikely to have significant
impact on the investigation of the velocity structures.
Systemic velocities derived from the HCN line are also consistent with those
derived from the CS line, as well as those from millimeter lines of
N$_{2}$H$^{+}$ (1--0) (\cite{mar97}) and H$^{13}$CO$^{+}$ (3--2) (\cite{gre97}).

\begin{figure}
  \begin{center}
    \FigureFile(120mm,120mm){figure1.eps}
  \end{center}
  \caption{HCN (4--3) (upper) and CS (7--6) (lower) spectra toward the
protostellar positions of L483 (left), B335 (middle), and L723 (right), taken with ASTE.}\label{fig:f1}
\end{figure}

\begin{table}
  \caption{Gaussian-Fitted Line Parameters of the Observed Submillimeter Emission}\label{tab:lines}
  \begin{center}
    \begin{tabular}{lccccccc}
      \hline\hline
            &\multicolumn{3}{c}{HCN ($J$=4--3)} &  &\multicolumn{3}{c}{CS ($J$=7--6)} \\
      \hline
      Source &T$_{B}$ &V$_{LSR}$     &$\Delta v$    &  &T$_{B}$ &V$_{LSR}$     &$\Delta v$     \\
             &(K)      &(km s$^{-1}$) &(km s$^{-1}$) &  &(K)      &(km s$^{-1}$) &(km s$^{-1}$) \\
       \hline
      L483 Center       &0.84 &5.62  &2.09  & &1.80 &5.52  &1.36 \\
      L483 -20$\arcsec$\footnotemark[$*$]   &0.51 &5.59  &1.40  &  &0.40 &5.94  &1.07   \\
      L483 average\footnotemark[$\dagger$]  &0.57 &5.64  &1.90  & &0.82 &5.59  &1.24  \\
      B335 Center       &0.58 &8.06  &0.69  &  &1.20 &8.18  &1.49   \\
      L723 Center       &0.31 &11.02 &3.05  &  &0.79 &11.25 &1.52   \\
       \hline
       \multicolumn{8}{@{}l@{}}{\hbox to 0pt{\parbox{85mm}{\footnotesize 
       \par\noindent
       \footnotemark[$*$] Position at the 20$\arcsec$ west of the protostar.
       \par\noindent
       \footnotemark[$\dagger$] Averaged over the central 20$\arcsec$ $\times$ 30$\arcsec$ region, 
where the central protostar locates at 10$\arcsec$ west and 10$\arcsec$ north from the bottom-left corner of the region.
     }\hss}}
    \end{tabular}

  \end{center}
\end{table}

\subsection{Spatial and Velocity Structures of the Submillimeter HCN and CS Emission}
\subsubsection{L483}

In Figure \ref{fig:f2}, we show total integrated intensity maps of
the HCN (4--3) and CS (7--6) emission in L483, along with the line profile
maps of the HCN (4--3) and CS (7--6) emission in Figure \ref{fig:f3}
and \ref{fig:f4}, respectively.
There appears a western extension both in the HCN and CS emissions,
which is consistent with the emission distribution in the 450 $\mu$m
continuum map (\cite{jor04}).
The intensity distribution both in the HCN and CS emission
is clearly different
from that of the circular beam shown at the bottom right corner of each panel
in Figure \ref{fig:f2},
which suggests that the structures traced by these submillimeter lines are
resolved with the present observations in L483.
The deconvolved size of the HCN emission
measured from a 2-dimensional Gaussian fitting to the image is
5500 $\times$ 3700 (AU) (P.A. = 78$^{\circ}$), while in the CS emission
only the major axis is resolved ($\sim$ 2300 AU).

\begin{figure}
  \begin{center}
    \FigureFile(120mm,120mm){figure2.eps}
  \end{center}
  \caption{Total integrated intensity maps (integrated velocity range 4.2 - 6.9 km s$^{-1}$)
of the HCN (4--3) (left) and CS (7--6) (right) emission in L483. Contour levels are
2, 4, 6 $\sigma$, and then 10 $\sigma$ in steps of 4 $\sigma$ (1 $\sigma$ = 0.0733 K km s$^{-1}$).
The highest contour in the HCN map is 18 $\sigma$ and that in the CS map 34 $\sigma$.
Crosses indicate observed positions, and open circles at the bottom right corner
beam sizes. Red and blue arrows show the direction of the redshifted and blueshifted 
molecular outflow, respectively, and the roots of the arrows indicate the
protostellar position.}\label{fig:f2}
\end{figure}

\begin{figure}
  \begin{center}
    \FigureFile(150mm,150mm){figure3.eps}
  \end{center}
  \caption{Line profile map of the HCN (4--3) emission in L483. Positions of each spectrum
are shown as crosses in Figure 2. Solid horizontal lines show zero levels,
and dashed vertical lines systemic velocity of 5.6 km s$^{-1}$. 3-channel bindings
are performed to increase the signal-to-noise ratio of the spectra.}\label{fig:f3}
\end{figure}

\begin{figure}
  \begin{center}
    \FigureFile(150mm,150mm){figure4.eps}
  \end{center}
  \caption{Same as Figure 3 but for the CS (7--6) emission.}\label{fig:f4}
\end{figure}

\begin{figure}
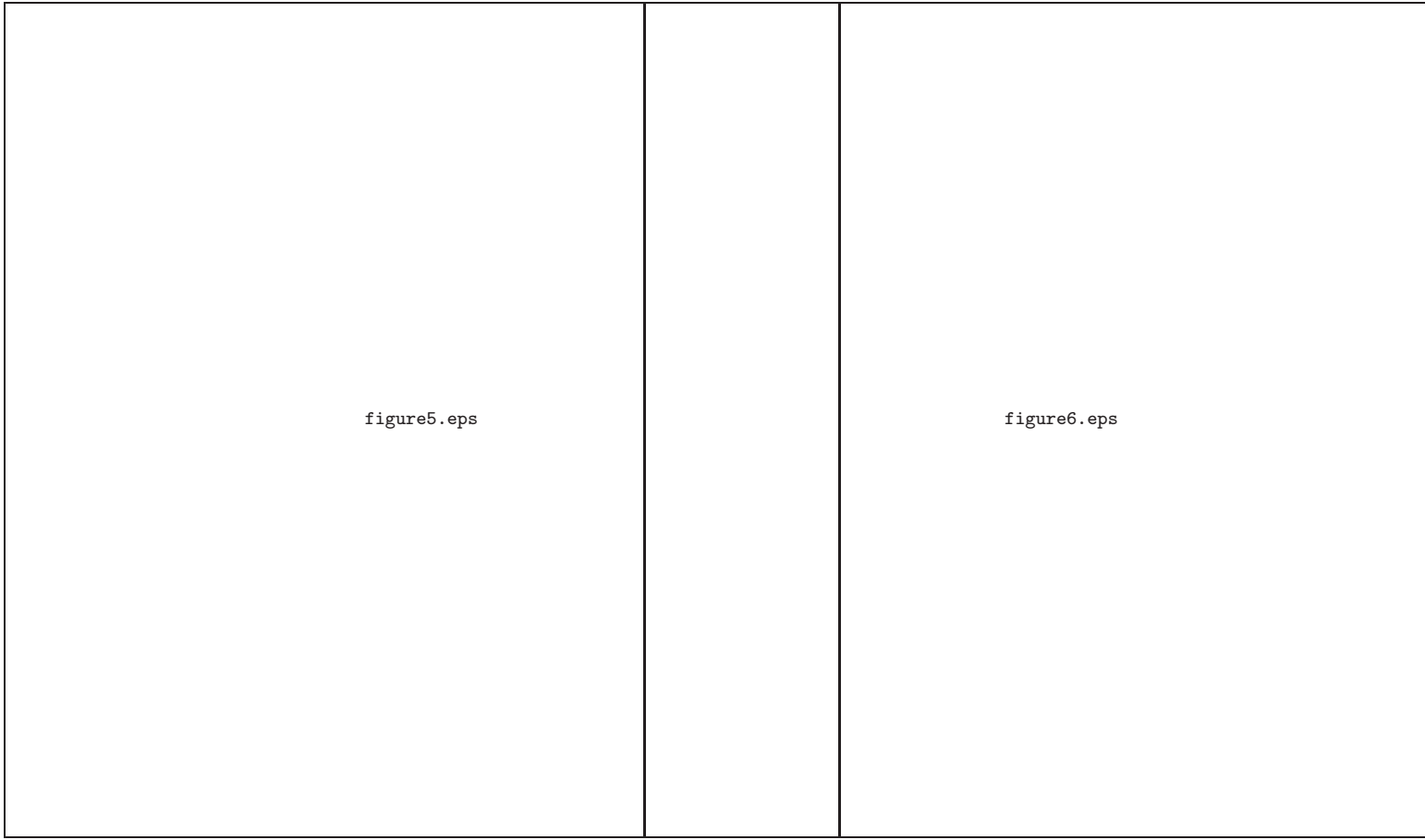

  \begin{center}
    \FigureFile(120mm,120mm){figure5.eps}
  \end{center}
  \caption{Position-velocity (P-V) diagrams of the CS (left) and HCN (right)
line across (upper) and along (lower) the axis of the associated molecular outflow
passing through the central stellar position in L483. Contour levels are
from 2 $\sigma$ in steps of 2 $\sigma$ (1 $\sigma$ = 0.133 K).
Dashed lines delineate detected velocity gradients.}\label{fig:f5}
\end{figure}

In Figure \ref{fig:f5}, we show Position-Velocity (P-V)
diagrams passing through the central stellar position
perpendicular and parallel to the axis of the associated molecular
outflow (E-W; see Figure \ref{fig:f2}) (\cite{buc99,hat99,taf00}).
There appears velocity gradients in the CS and HCN emission
both across and along the direction of the associated outflow.
Perpendicular to the outflow, the northern part is blueshifted while
the southern part is redshifted, and along the outflow axis the eastern
part appears blueshifted and the western part redshifted.
Although the loci of the velocity gradients shown in Figure \ref{fig:f5}
are somewhat arbitrary,
these trends are also evident in the line profile maps of Figure \ref{fig:f3}
and \ref{fig:f4}.
Across the outflow axis, similar velocity gradients
are also seen in the 3-mm counterpart of the
CS (2--1) line (\cite{jor04}) and in the 3-mm C$_{3}$H$_{2}$
(2$_{12}$--1$_{01}$) line (\cite{par00}).
The amount of
the velocity gradient observed in the submillimeter CS and HCN lines
is measured to be $\sim$ 230 km s$^{-1}$ pc$^{-1}$, which is much larger
than that in the 3-mm C$_{3}$H$_{2}$ line ($\sim$ 10 km s$^{-1}$ pc$^{-1}$)
even though the spatial resolution of their
3-mm observations (= 14$\arcsec$ $\times$ 10$\arcsec$) is slightly 
better than ours (\cite{par00}).
At higher spatial resolution ($\sim$ 7$\arcsec$), the velocity gradient
in the 3-mm counterpart of the CS (2--1) line is measured to be
$\sim$ 410 km s$^{-1}$ pc$^{-1}$ (\cite{jor04}), twice higher than that obtained by
our lower-resolution submillimeter observations. These velocity gradients
most likely trace rotational gas motion around the protostar at different radii.

Along the outflow axis,
the velocity gradient is measured to be $\sim$ 10 km s$^{-1}$ pc$^{-1}$ in these
submillimeter lines. Interestingly, in the 3-mm counterpart of the CS (2--1) line and
other 3-mm lines such as N$_{2}$H$^{+}$ (1--0) the eastern part shows redshifted emission
and vice versa, and the velocity gradient is measured to be
$\sim$ 280 km s$^{-1}$ pc$^{-1}$ (\cite{jor04}). Hence,
the velocity gradient traced by the submillimeter lines
is opposite to that traced by the millimeter lines.
The associated molecular outflow observed in CO (2--1; 4--3) lines
shows redshifted emission at the east of the protostar and blueshifted
emission as well as the K$'$ reflection nebula at the west (\cite{hat99,taf00}).
Therefore, the velocity gradient in the submillimeter lines
is also opposite to that of the associated outflow.
In L483, the E-W velocity gradient traced by millimeter lines and
CO lines is interpreted as a motion of the outflowing gas
and the entrained dense gas (\cite{ful00,taf00,jor04}).
The different velocity gradient found in our submillimeter observations
suggest that the submillimeter emissions trace a different gas component.

\subsubsection{B335}

Figure \ref{fig:f6}, Figure \ref{fig:f7} and \ref{fig:f8} show total integrated intensity maps 
and line profiles maps of
the HCN (4--3) and CS (7--6) emission in B335. In B335, we
could not spatially resolve the structure traced by these submillimeter lines.
The CS line intensity, as well as the CS line width, is much larger than
those of the HCN line.
Figure \ref{fig:f9} shows P-V diagrams passing through the central stellar position
across and along the axis of the the associated molecular outflow in B335
(E-W; see Figure \ref{fig:f6}) (\cite{hir92,cha93}). In the CS emission,
there appears velocity gradients both across and along the direction of
the associated outflow, whereas the HCN emission is too weak to trace the
velocity structure.
Perpendicular to the outflow, the northern part is blueshifted while
the southern part is redshifted, and along the outflow axis
the western part appears to be blueshifted and the eastern part
redshifted. These velocity gradients are also seen in the line profile map
of Figure \ref{fig:f8}.

Across the outflow axis,
the velocity gradient traced by the submillimeter CS line ($\sim$ 160 km s$^{-1}$ pc$^{-1}$)
has an opposite sense to that traced by the millimeter
H$^{13}$CO$^{+}$ (1--0) line, that is, in the H$^{13}$CO$^{+}$ line
the northern part is redshifted and the southern part blueshifted
($\sim$ 50 km s$^{-1}$ pc$^{-1}$) (\cite{sai99}).
The same trend is
also seen along the outflow axis. In the submillimeter CS emission
the western part is blueshifted and the eastern part
redshifted ($\sim$ 110 km s$^{-1}$ pc$^{-1}$), while in the millimeter 
H$^{13}$CO$^{+}$ emission the western part is redshifted and vice versa
($\sim$ 780 km s$^{-1}$ pc$^{-1}$) (\cite{sai99}).
Aperture synthesis images in the optically thick $^{13}$CO (1--0) line
show well-separated blue- and redshifted outflow lobes at the east and
west of the protostar of B335, respectively (\cite{cha93}).
Hence the velocity gradient revealed by our submillimeter observations
is also opposite to that of the associated outflow.
The E-W velocity gradient traced by the millimeter H$^{13}$CO$^{+}$
line is interpreted as an infalling gas motion in the flattened
disklike envelope surrounding the protostar, perpendicularly to the
axis of the associated molecular outflow (\cite{sai99}).
The different velocity gradient in the submillimeter CS line
implies the different origin of the gas motion from the infalling gas.

\begin{figure}
  \begin{center}
    \FigureFile(120mm,120mm){figure6.eps}
  \end{center}
  \caption{Total integrated intensity maps (integrated velocity range 6.8 - 9.4 km s$^{-1}$
in the CS emission and 7.7 - 8.3 km s$^{-1}$ in the HCN emission)
of the HCN (4--3) (left) and CS (7--6) (right) emission in B335.
Contour levels are
2, 4, 6 $\sigma$, and then 10 $\sigma$ in steps of 4 $\sigma$ (1 $\sigma$ = 0.0733 K km s$^{-1}$).
The highest contour in the CS map is 22 $\sigma$.
Crosses indicate observed positions, and open circles at the bottom right corner
beam sizes. Red and blue arrows show the direction of the redshifted and blueshifted
molecular outflow, respectively, and the roots of the arrows indicate the
protostellar position.}\label{fig:f6}
\end{figure}

\begin{figure}
  \begin{center}
    \FigureFile(150mm,150mm){figure7.eps}
  \end{center}
  \caption{Line profile map of the HCN (4--3) emission in B335. Positions of each spectrum
are shown as crosses in Figure 6. Solid horizontal lines show zero levels,
and dashed vertical lines systemic velocity of 8.1 km s$^{-1}$. 3-channel bindings 
are performed to increase the signal-to-noise ratio of the spectra.}\label{fig:f7}
\end{figure}

\begin{figure}
  \begin{center}
    \FigureFile(150mm,150mm){figure8.eps}
  \end{center}
  \caption{Same as Figure 7 but for the CS (7--6) emission.}\label{fig:f8}
\end{figure}

\begin{figure}
  \begin{center}
    \FigureFile(120mm,120mm){figure9.eps}
  \end{center}
  \caption{P-V diagrams of the CS (left) and HCN (right)
line across (upper) and along (lower) the axis of the associated molecular outflow
passing through the central stellar position in B335. Contour levels are
from 2 $\sigma$ in steps of 2 $\sigma$ (1 $\sigma$ = 0.133 K).
Dashed lines delineate detected velocity gradients.}\label{fig:f9}
\end{figure}

These ASTE results in L483 and B335 may suggest that the millimeter and submillimeter
lines trace different gas components with different kinematics in the low-mass
protostellar envelopes. On the other hand, we need further observational confirmation to
verify the presence of the different velocity structure, since the spatial resolution
as well as the calibration accuracy of the present submillimeter observations are limited.
High-resolution imaging observations with ALMA are crucial to unambiguously
clarify the velocity structure and the origin of the submillimeter emissions
in the low-mass protostellar envelopes.

\section{Discussion}
\subsection{Physical Conditions of Molecular Gas traced by the Submillimeter Lines\\
in the Low-mass Protostellar Envelopes}

Our ASTE observations have resolved the spatial and/or kinematical structures
of the submillimeter HCN and CS emission in L483 and B335.
The intensities and extents
should trace structures of gas densities, temperatures, and molecular abundances in
these envelopes. In order to investigate these physical
conditions of the protostellar envelopes traced by the extended submillimeter
emissions, we performed statistical equilibrium calculations of the
submillimeter lines based on the Large Velocity Gradient (LVG) model (\cite{gol74,sco74}).

For the model calculations, we employed values of the dipole moment, rotational and centrifugal
constants of the HCN and CS molecule listed by Winnewisser et al. (1979).
Collisional transition rates of HCN were taken from Green and Thaddeus (1974)
and those of CS from Green and Chapman (1978).
Rotational energy levels included in our calculations are up to
$J$ = 8 (153 K) for HCN and $J$ = 13 (214 K) for CS.
These energy levels are high enough to discuss physical conditions of molecular
gas in low-mass protostellar envelopes (\cite{mor95}).

In Figure 10, we show results of our calculations. The HCN and CS line intensities
increase monotonically as the gas density ($\equiv$ $n_{H2}$),
molecular abundance per unit velocity gradient ($\equiv$ $X/dv/dr$),
or the gas kinetic temperature ($\equiv$ $T_{k}$) increases.
Since we have only one transition of HCN and CS, we cannot fully constrain
the physical conditions traced by these submillimeter lines, but
we can still make some arguments as follows.
In L483, the HCN and CS abundance is estimated to be
$X_{HCN}$ = 2.0 $\times$ 10$^{-9}$ and $X_{CS}$ = 6.8 $\times$ 10$^{-10}$,
respectively (\cite{joe04}). The observed HCN and CS line width ($\equiv$ $dv$)
at the center of L483 is $\sim$ 2.1 km s$^{-1}$ and 1.4 km s$^{-1}$ ($\S$3.1)
and the deconvolved size ($\equiv$ $dr$)
along the major axis of the HCN and CS emission is 0.027 pc and 0.011 pc ($\S$3.2).
Hence, the $X/dv/dr$ value can be derived to be $\sim$ 2.6 $\times$ 10$^{-11}$
for HCN and $\sim$ 5.6 $\times$ 10$^{-12}$ for CS.
Then, at the 20$\arcsec$ = 4000 AU west from the central protostar of L483
where the HCN and CS lines are detected to be 0.51 (K) and 0.40 (K), respectively (Table 2),
the value of $n_{H2}$ must be higher than $\sim$ 6 $\times$ 10$^{6}$ cm$^{-3}$ if
$T_{k}$ = 10 (K), or the value of $T_{k}$
must be higher than $\sim$ 40 (K) if $n_{H2}$ = 6 $\times$ 10$^{5}$ cm$^{-3}$.
At the central protostellar positions of L483
and B335, the same $X/dv/dr$ values provide
the estimate of $n_{H2}$ $>$ 10$^{7}$ cm$^{-3}$ if $T_{k}$ $\sim$ 10 (K), or the estimate of
$T_{k}$ $>$ 40 (K) if $n_{H2}$ $\sim$ 10$^{6}$ cm$^{-3}$, averaged over the 22$\arcsec$ ASTE beam.
These arguments imply that high-density
($\gtrsim$ 6 $\times$ 10$^{6}$ cm$^{-3}$) or high-temperature ($\gtrsim$ 40 K) molecular gas
could be more extended than $\sim$ 2000 AU in radius, particularly evident at the
4000 AU west from the protostar of L483.

The presence of such an extended high-density or high-temperature gas in low-mass
protostellar envelopes
is also supported by other recent submillimeter-line observations
of low-mass protostellar envelopes. Takakuwa et al. (2006) have made single-dish
and interferometric observations of IRAS 16293-2422 in the HCN (4--3) line with
the JCMT and the SMA and have found that there is an extended ($>$ 3000 AU) HCN
emission plus compact ($\sim$ 500 AU) disklike structure associated with the
protostar. CS (7--6) observations of L1551 IRS5 with the SMA
imply that the submillimeter CS emission is more extended than $\sim$ 1500 AU
(\cite{tak04}). However, it is still less clear why these submillimeter
molecular lines which trace high-density or high-temperature gas
can be so extended.
Masunaga et al. (1998, 2000) have constructed a one-dimensional
radiation hydrodynamic model in the formation of an 1 M$_{\odot}$ protostar.
Their model suggests that during the main accretion phase
the $T_{k}$ and $n_{H2}$ values at the radius of 4000 AU (i.e., L483 20$\arcsec$ west)
are $\sim$ 15 K and $\sim$ 2 $\times$ 10$^{5}$ cm$^{-3}$,
and at the radius of 2000 AU $\sim$ 20 (K) and $\sim$ 10$^{6}$ cm$^{-3}$,
respectively. These theoretically-predicted values seem to be slightly lower
to explain the observed submillimeter emissions, although the limited
spatial resolution as well as the only single transitions of our ASTE data
prevents us from making direct comparison between the observational and theoretical results.
On the other hand, the western extension of the submillimeter emission in L483
seems to be aligned with the western component of the associated outflow (\cite{taf00,jor04}),
and there might be another heating mechanism such as shocks due to the interaction between the
outflow and the ambient gas (\cite{ume92,hir01}).

As described in $\S$3.2 the velocity structures traced by the submillimeter
lines are different from those traced by millimeter lines both in L483 and B335,
which suggests that the submillimeter emissions have a different origin
and that the spherically-symmetric radiation hydrodynamic model is not adequate to fully
explain our observational results. We will discuss on these different velocity structures
in the next section.

\begin{figure}
  \begin{center}
    \FigureFile(150mm,150mm){figure10.eps}
  \end{center}
  \caption{Results of our LVG model calculations. Brightness temperatures of the CS (7--6) (left)
and HCN (4--3) (right) lines are plotted in contours as a function of the gas density
($\equiv$ $n_{H2}$) and the molecular abundance per unit velocity gradient
($\equiv$ $X/dv/dr$) at each gas kinetic temperature ($T_{k}$ = 10, 20, 40, 60, and 100 K
as shown in the Figure). Contour levels are shown in the Figure, and the
lowest five contours correspond to the observed line intensities at the representative
positions listed in Table 2.}\label{fig:f10}
\end{figure}

\subsection{Different Velocity Structures traced by the Submillimeter Lines\\
in the Low-mass Protostellar Envelopes}

Our submillimeter-line observations of the low-mass protostellar envelopes
have revealed that along the axis of the associated molecular outflows
the sense of the velocity gradient traced by the submillimeter lines
is opposite to that traced by the previous millimeter observations or the molecular outflows
themselves ($\S$3.2).
In L483, millimeter CS (2--1) and N$_{2}$H$^{+}$ (1--0) lines as well as the associated CO outflow
show redshifted emissions at the east of the protostar and blueshifted emissions at the west
(\cite{hat99,jor04}),
while the submillimeter CS (7--6) and HCN (4--3) lines exhibit blueshifted emissions
at the east and redshifted emissions at the west of the protostar.
In B335, the associated CO outflow and the millimeter C$^{18}$O (1--0) and
H$^{13}$CO$^{+}$ (1--0) lines show blueshifted emissions at the east of the protostar
and redshifted emissions at the west (\cite{cha93,sai99}), whereas the submillimeter CS and
HCN lines show an opposite velocity gradient along the same axis.
In this subsection, the origin of these
different velocity structures traced by the submillimeter lines is discussed.

One possible interpretation of the different velocity structure in the submillimeter lines
is that the submillimeter lines trace the other side of the conical shell of the outflow
than that traced by the millimeter lines incidentally, whose axis is close to the plane of the sky.
The inclination angle of the associated outflow from the plan of the sky
in L483 is estimated to be $\sim$ 45$^{\circ}$
and there appears a weak redshifted outflow component at the western side (\cite{par00,taf00}),
whereas the inclination angle in B335 is derived to be $\sim$ 10$^{\circ}$ and
there are also redshifted and blueshifted outflow components at the east and west of the protostar,
respectively (\cite{hir92}).
Although the main outflow component of the blueshifted and redshifted gas locates
at the west and east of the central source of L483 (\cite{hat99}) and and at the east and west of
B335 (\cite{cha93}), it is possible that
the observed submillimeter emission trace those other outflow
components incidentally. In such a case, the velocity gradient in the submillimeter emission
shows the opposite sense to that of the main velocity gradient traced by the millimeter lines.
The observed elongation of the submillimeter emission along the outflow axis in L483 favors
this interpretation, and the higher-temperature ($\gtrsim$ 40 K) gas at the western extension
in L483 discussed in the previous subsection may reflect the shock interaction between
the outflow and the ambient dense gas (\cite{ume92,hir01}).


Another possible and more interesting interpretation of the observed velocity structures
in the submillimeter lines is
an expanding gas motion which is perpendicular to the outflow axis, as shown in Figure 11.
This figure demonstrates that the submillimeter lines trace molecular gas at the surface
of the flattened disklike envelope, which is being stripped away by some mechanism such 
as the stellar wind,
while the millimeter lines trace infalling gas at the midplane of the envelope.
Here, on the side of the blueshifted outflow the expanding gas is observed as a redshifted
component and on the side of the redshifted outflow as a blueshifted component,
while the infalling gas components show the same velocity gradient as that of the outflow.
Hence, this configuration can explain
the different velocity gradient observed in the submillimeter lines
from that of the outflow components or infalling gas traced by the millimeter lines.

Similar velocity structures along the perpendicular direction to the polar outflow
have also been found in near-infrared observations of young stellar objects.
Pyo et al. (2005) have observed lower-velocity [Fe II] winds with a wide opening angle of
$\sim$ 100$^{\circ}$ in L1551 IRS5, as well as well-collimated higher-velocity jets
along the polar axis. Comparable results were also obtained for the young T Tauri star DG Tau
(\cite{pyo03}). They suggest that
such a wind with a wide opening angle will be effective in sweeping up
envelope material from the close vicinity of its driving source. The
observed velocity structures in the submillimeter lines may trace such a
gas motion in the envelope, although the spatial scale in the submillimeter results
($>$ 2000 AU) is much larger than that of the infrared observations ($\sim$ 100 - 400 AU).
In L483, the line core emission in the HCO$^{+}$ (3--2) line does show the same velocity
gradient as our submillimeter results, while the HCO$^{+}$ (3--2) line wing emission
traces the bipolar outflow itself (see Figure 15 in \cite{gre97}). 
At the surface of the flattened disklike envelopes molecular gas is irradiated 
by the stellar photons and hence the gas temperature is likely to be higher than 
that at the midplane. Then, the warmer surface regions are selectively traced 
by the submillimeter observations whereas the colder, infalling regions at the midplane are traced 
by the millimeter lines alone. In fact, Spaans et al. (1995) proposed that
the narrow $^{12}$CO and $^{13}$CO $J$=6--5 emission observed toward many
low-mass young stellar objects is produced at the surface of the 
circumstellar envelope surrounding the cavity evacuated by the bipolar outflow,
heated by the radiation field generated in the inner
part of the accretion disk. Recent two-dimensional radiative transfer models in
protostellar envelopes have also revealed that at the surface of the
flattened disklike envelope the gas temperature becomes higher than 
that predicted from spherically symmetric models (\cite{nak03,whi03}).
The observed submillimeter HCN and CS emission may have such an origin.

The interpretation of the expanding flattened envelope, however, has one weak point,
that is, in L483 the observed submillimeter emissions are elongated $along$ the outflow direction.
If the submillimeter lines trace expanding gas motion at the surface of the disklike envelope,
the elongation should be perpendicular to the axis of the associated outflow. In order
to compensate for this contradiction, non polar-symmetric geometry
and/or the flared shape of the envelope must be incorporated. With the present data,
we cannot determine
which scenario, the incidental projection effect or the expanding flattened envelope
with the flared and/or non polar-symmetric shape, is more plausible.

On the other hand, we have also found an
opposite velocity gradient in the submillimeter lines to that in the millimeter lines
across the outflow axis in B335. The origin of the difference is less clear,
but the expanding gas traced by the submillimeter lines may have a different
rotational motion.
In any sense, our new submillimeter mapping observations have presented us an implication
that submillimeter lines trace different kinematics from that traced 
by millimeter lines, and that there is another velocity structure which
previous millimeter observations could not find in low-mass protostellar envelopes.

\begin{figure}
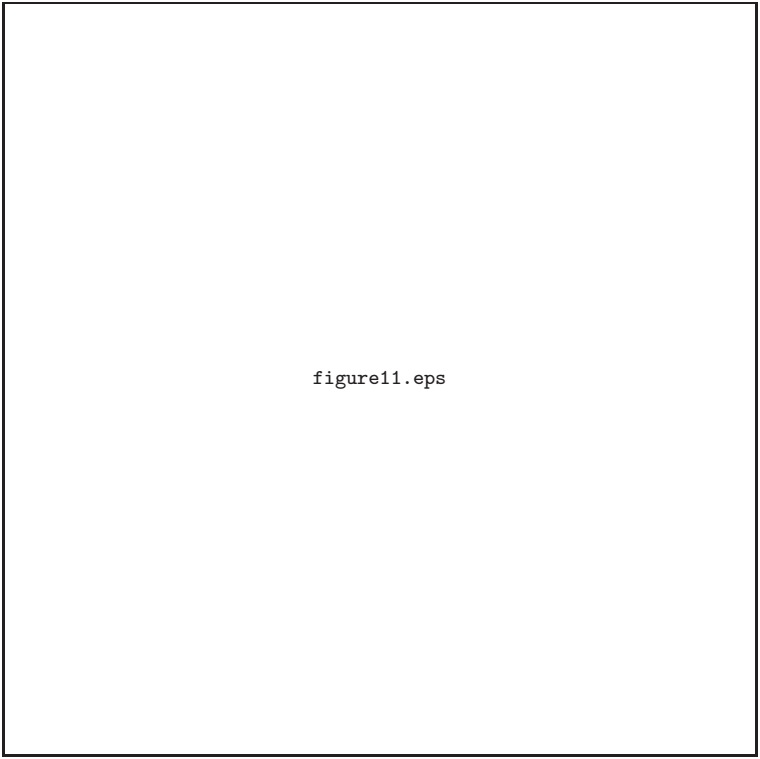

  \begin{center}
    \FigureFile(100mm,100mm){figure11.eps}
  \end{center}
  \caption{A schematic diagram to interpret the observed velocity gradient
of the submillimeter emission in the low-mass protostellar envelopes.
Observers' line of sight is normal to the paper.
The bold black ellipse represents a flattened disklike envelope.
This disklike envelope is tilted, and the near side is to the left.
The associated blueshifted and redshifted molecular outflows perpendicular to the disklike
envelope are shown in light blue and red colors, respectively, and the blueshifted outflow
locates at the lefthand, near side. At the midplane of the disklike envelope there is an infalling
gas motion as described in the black arrows. At the surface of the disklike envelope
there is an expanding gas motion as shown in red and blue arrows, which is
traced by the present submillimeter observations. Here, the left-half
portion is redshifted while the other half blueshifted, and this velocity structure
has an opposite sense to that of the outflow or the infalling gas motion.
The configuration in this diagram represents the case of B335. For the case of
L483 one can just flip this configuration horizontally.}\label{fig:f11}
\end{figure}

\section{Summary}

We have made mapping observations of low-mass protostellar envelopes of L483 and
B335 as well as a single-point observation of L723
in the HCN ($J$=4--3) and CS ($J$=7--6) lines with ASTE.
Main results of our new submillimeter observations are summarized
as follows.

\begin{itemize}
\item[1]
We have detected both the HCN and CS lines toward all the sources.
Typical intensities and line widths are 0.5 K and 2.0 km s$^{-1}$ in the HCN line
and 1.0 K and 1.5 km s$^{-1}$ in the CS line, respectively.
The CS-to-HCN intensity ratio is similar across the three samples (2.1 - 2.5),
and there may be a correlation between the source luminosity and the intensity
of the submillimeter lines.

\item[2]
Mapping observations of L483 in the HCN and CS lines show that
the submillimeter emissions are extended to the west of the protostar,
consistent with the distribution of the dust continuum emission, and
that the deconvolved size is $\sim$ 5500 AU $\times$ 3700 AU
(P.A. = 78$^{\circ}$) in the HCN emission.
The extent of the submillimeter emissions in L483 implies the presence
of higher-temperature ($\gtrsim$ 40 K) or higher-density
($\gtrsim$ 6 $\times$ 10$^{6}$ cm$^{-3}$) gas at
4000 AU away from the central protostar. The existence of such high-temperature or
high-density gas at 4000 AU away from the protostar is unlikely to be explained
via heating from the central protostar with the spherically symmetric geometry in
the envelope.

\item[3]
Both in L483 and B335, we found velocity gradients in the submillimeter lines
along and across the axis of the associated molecular outflow.
The sense of the velocity gradient
traced by the submillimeter lines along the outflow direction
is opposite to that of the results from the previous millimeter observations or
the associated molecular outflow itself, both in L483 and B335.
One possible interpretation of the different velocity gradient is that the
submillimeter lines trace the other side of the conical shape of the outflow
incidentally, where the outflow axis is close to the plane of the sky.
Alternatively, expanding
gas motions at the surface of the flattened disklike envelopes, or
gas motions to strip off the envelope, irradiated from the central star directly, 
can explain the observed submillimeter velocity structure.
\end{itemize}

We are grateful to R. Kawabe and N. Ohashi for their fruitful discussions.
We thank all the ASTE staff for their dedicated support of the telescope
and observatory operations. We acknowledge the anonymous referee, whose
comments polish up the manuscript significantly.

\end{document}